\documentclass{iitpressproc}

\usepackage{graphicx}
\usepackage{amsmath}
\usepackage{url}

\begin{document}

\title{Parton Distributions and Event Generators}

\author{Stefano Carrazza, Stefano Forte
\address{ Dipartimento di Fisica, Universit\`a di Milano and
INFN, Sezione di Milano, Via Celoria 16, I-20133 Milano, Italy}\\[2ex]
 Juan Rojo
\address{ PH Department, TH Unit, CERN, CH-1211 Geneva 23, Switzerland}
}

\maketitle

\begin{abstract}
We present  the implementation within  the {\tt Pythia8} event generator of a set of
parton distributions based on NNPDF methodology.
We construct a set of leading-order parton distributions with QED
corrections,  NNPDF2.3QED LO set, based on the same data as the
previous NNPDF2.3 NLO and NNLO PDF sets. We compare this PDF set to
its higher-order counterparts, we discuss its implementation as an
internal set in {\tt Pythia8},  and we use it to study
some of the phenomenological implications of photon-initiated contributions
for dilepton production at hadron colliders.
\end{abstract}

\section{PDFs and event generators}

The needs of physics at the LHC require an increasingly accurate
control of the parton substructure of the nucleon: for example, this is
a necessary ingredient in the
accurate determination of Higgs couplings~\cite{petriello,rebuzzi}, 
which  in turn is essential both for precision
 determination of Standard Model parameters and for indirect searches for
New Physics. 
Current sets of parton distributions~\cite{Forte:2013wc} 
are based on
increasingly refined theory, use an increasingly wide dataset (now
also extended to LHC data) and attempt to arrive at an estimation of
uncertainties which is as reliable as possible. An important
ingredient in achieving all of these goals is the integration of
parton distributions within Monte Carlo event
generators~\cite{Buckley:2011ms}. Indeed, 
parton showering and hadronization are necessary in order to bridge
perturbative QCD calculation with the quantities which  are actually
measured in experiments, all the more so as  less inclusive
observables are considered,  even though also for observable which are in
principle inclusive (such as the production of gauge bosons)
comparisons are best made between theoretical predictions, and data
collected in an experimental fiducial region. 

Whereas next-to-leading (NLO) order Monte Carlo tools play an increasingly
important role, leading-order (LO) Monte Carlo simulations are still
commonly used in a variety of applications.
 Furthermore, both LO and NLO Monte Carlo
event generators typically rely on leading-order PDFs for the description of
multiple-parton interactions and the underlying event, so that in fact
generators which include a hadronization model,  such as {\tt Pythia} (and
specifically its current version,  {\tt
  Pythia8}~\cite{Sjostrand:2007gs}) are tuned using one or more
`native' PDF sets 

In this short contribution, we will discuss the NNPDF2.3QED LO PDF
set, and its implementation within {\tt
  Pythia8}: this is a PDF set which is based on  the successful NNPDF
methodology, which strives to minimize
theoretical bias and construct statistically reliable parton
distributions,
recently used to produce a first global set of PDFs using
LHC data, NNPDF2.3~\cite{Ball:2012cx}.
This PDF set was
 subsequently used to construct
a first set of PDFs with QED corrections and a photon distribution
determined by experimental data, NNPDF2.3QED~\cite{Ball:2013hta}. Recent general
reviews of parton distributions are presented in
Refs.~\cite{DeRoeck:2011na,Perez:2012um,Forte:2013wc}.

\section{The NNPDF2.3QED~LO parton set}

The NLO and NNLO NNPDF2.3QED PDF sets were recently presented
in~\cite{Ball:2013hta}.
In these sets, the evolution of quark and gluon PDFs is consistently
performed using  combined QCD$\otimes$QED evolution equations, and the
photon PDF $\gamma(x,Q^2)$ is determined from LHC vector boson
production and deep-inelastic scattering data.
In order to construct a corresponding leading-order set, NNPDF2.3QED~LO,  we start from the  NNPDF2.1LO PDF
sets~\cite{Ball:2011uy}, with two different values of $\alpha_s(M_Z)$:
0.119 and 0.130. Note that in Ref.~\cite{Ball:2011uy} further LO sets were
constructed, NNPDF2.1 LO* in which the  momentum sum
  rule was not imposed; however, this choice, sometimes advocated, did
  not turn out to be especially advantageous, hence we will not
  discuss these sets further.

The NNPDF2.3QED~LO set is constructed by combining the PDFs from the
NNPDF2.1~LO set with the photon PDF from the 
NNPDF2.3QED NLO set at $Q_0^2=2$
GeV$^2$, and then evolving upwards this boundary condition with combined LO
QCD$\otimes$QED evolution equations, including $O(\alpha_s)$ and $O(\alpha)$,
but not $O(\alpha\alpha_s)$ terms.
This procedure (which clearly retains LO
QED+QCD accuracy) is justified because of the very mild
correlation between the photon and the other PDFs, and the large
uncertainty on the photon PDF itself~\cite{Ball:2013hta}.
The set of PDFs thus obtained is then also evolved down to
$Q^2=1$ GeV$^2$: whereas leading-twist perturbative QCD might not be
accurate in this region, low-scale PDFs are necessary for tunes
of the underlying event and minimum bias physics in shower Monte
Carlos.

The combined QCD$\otimes$QED evolution has been performed with the {\tt APFEL}
package~\cite{Bertone:2013vaa}. Among the various forms of the
solution to the evolution equation, differing by
$O(\alpha\alpha_s)$ terms which are beyond our accuracy, we use the
so-called
{\tt QECDS}~\cite{Bertone:2013vaa} solution, which
was also used for the construction of the NNPDF2.3QED NLO and NNLO sets.
Strict positivity of all  LO PDFs has been imposed in 
the relevant range of $x$ and $Q^2$.

\begin{figure}[h]
\centering
\includegraphics[width=0.75\columnwidth]{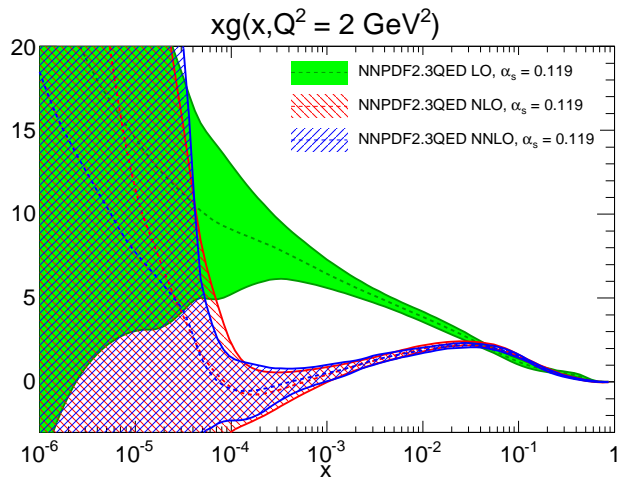}
\caption{\small The gluon PDF at LO, NLO and NNLO
in the NNPDF2.3QED sets.
\label{figGluon}}
\end{figure}

In Fig.~\ref{figGluon} we show the gluon PDF in the LO, NLO and NNLO
NNPDF2.3QED fits.
The much larger small-$x$ gluon is a well-known feature of LO PDF
sets, due to the need to  compensate for 
missing NLO terms  when fitting deep-inelastic structure function
data. It  is an important
ingredient for tunes of soft QCD dynamics at hadron colliders.

\begin{figure}
\centering
\includegraphics[width=0.75\columnwidth]{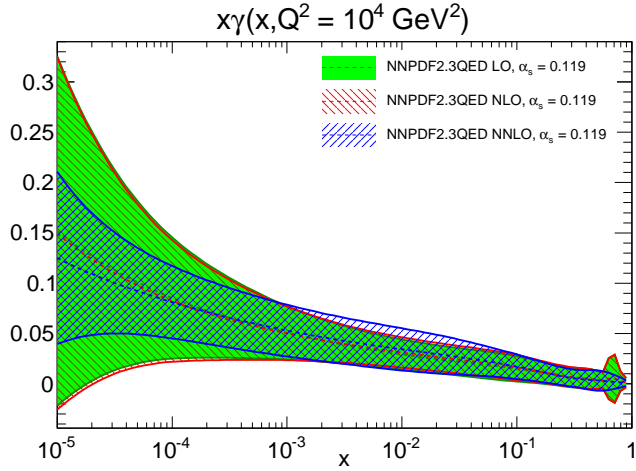}
\caption{\small The photon PDF at LO, NLO and NNLO
in the NNPDF2.3QED sets.
\label{figPhoton}}
\end{figure}

In Fig.~\ref{figPhoton} we also show the photon PDF at  
$Q^2=10^4$ GeV$^2$,  at LO, NLO and NNLO. 
The small differences seen arise due to  the different
evolution of quarks and gluons and their mixing with the
photon through evolution equations.

\begin{figure}[h]
\centering
\includegraphics[width=0.75\columnwidth]{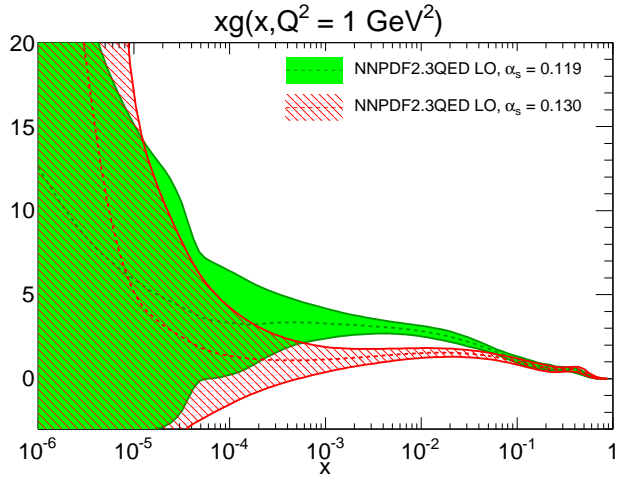}
\caption{\small The small-$x$ gluon PDFs in the NNPDF2.3QED~LO set for
   $\alpha_{s}=0.119$ and 0.130.
  \label{figGluon_1gev}}
\end{figure}

Finally, in Fig.~\ref{figGluon_1gev} we compare the gluon PDF in the
LO sets corresponding to the two different values of 
$\alpha_s(M_Z)=0.119$ and $0.130$: because of the slower running of
$\alpha_s$ at LO, the smaller value is more accurate at higher scale,
and the larger value at low scales. Reassuringly, in the small
$x \le
10^{-4}$, relevant for tunes of soft physics at
hadron colliders, the two sets turn out to agree within their large
uncertainties. 

\section{Implementation in {\tt Pythia8} and phenomenological implications}

The NNPDF2.3QED~LO sets, with two different $\alpha_s$ values, together
with their NLO and NNLO counterparts, have been implemented as
internal PDF sets in {\tt Pythia8}~\cite{Sjostrand:2007gs} starting with {\tt
  v8.180}, and there is ongoing work by the {\tt
  Pythia8} authors towards providing a complete new tune based on
NNPDF2.3QED~LO, including all the relevant constraints from LHC and
previous lower-energy colliders~\cite{sjospriv}.

For the time being, we will illustrate some of the phenomenological
implications of the NNPDF2.3QED~LO set by generating events with {\tt
  Pythia8} for processes in which photon-initiated contributions are
substantial.
As a case study, we consider dilepton production at the LHC~14TeV.
Related studies were presented in the original NNPDF2.3QED
paper~\cite{Ball:2013hta} but were restricted to the parton level, while
now we include the effects of  the initial state parton shower and
underlying event with the standard {\tt Pythia8} tune.
The QED shower option of {\tt Pythia8} is turned off.
We generate events for $q\bar{q} \to \gamma^*/Z \to l^+l^-$
and for $\gamma \gamma  \to l^+l^-$, and compare the relative contributions
of the two different initial states.
We consider both electrons and muons in the final state.

\begin{figure}[h]
\centering
\includegraphics[width=0.75\columnwidth]{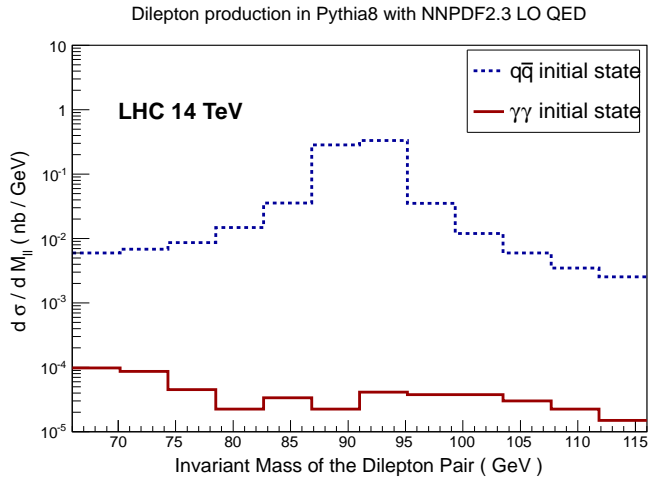}
\caption{\small Invariant mass distributions of the dilepton pair at
  the LHC 14 TeV, computed with NNPDF2.3QED LO and {\tt Pythia8}.
The contribution from the $q\bar{q}$ and $\gamma\gamma$ initiated
subprocess are separately shown.
No kinematical cuts have been applied.
\label{figPythia8}}
\end{figure}

The invariant mass distributions of the dilepton
pair at the LHC 14 TeV, without any kinematical cut, 
is shown in Fig.~\ref{figPythia8}, in the $Z$ peak mass
region.
We shown separately the contributions from the $q\bar{q}$ and
$\gamma\gamma$ initiated subprocesses (though experimentally they
cannot be separated, as they lead to the
same final state).
Is clear that in this region the $q\bar{q}$ contribution is much
larger, while the $\gamma\gamma$ contribution is at the permille level.
The total (leading order) 
cross section, including branching fractions, is found to be
around
3.2~nb, 
in agreement with MCFM when run with the same input PDF set.
We conclude that  photon-initiated contributions are generally not required
in the $Z$ peak region, except perhaps for high precision studies, such as
the determination of the $W$ boson mass, where a permille accuracy in the distributions
is required~\cite{Bozzi:2011ww}.

\begin{figure}[h]
\centering
\includegraphics[width=0.75\columnwidth]{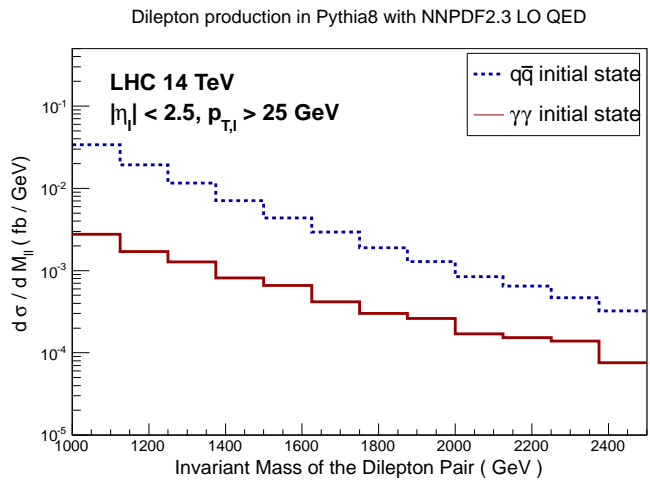}
\caption{\small Same as Fig.~\ref{figPythia8} but now in the high
dilepton mass region.
Realistic kinematical cuts have been applied to the events, see text.
 \label{figPythia82}}
\end{figure}

The situation is quite different if we consider the high-mass tail.
In Fig.~\ref{figPythia82} we show the region of dilepton
invariant masses $M_{ll}$ between
1 TeV and 2.5 TeV.
We have applied realistic kinematical cuts based on the typical
ATLAS and CMS acceptances, namely, we require $p_{T,l}\ge 25$~GeV and
$|\eta_{l}| \le 2.5$.
It is clear that now the photon-initiated contributions to the event
yields
are rather more significant, ranging from
10\% at low masses to up to 50\% at high masses.
Therefore, photon-induced contributions are an important background
for New Physics searches in electroweak production at
high invariant masses.

\begin{figure}[h]
\centering
\includegraphics[width=0.75\columnwidth]{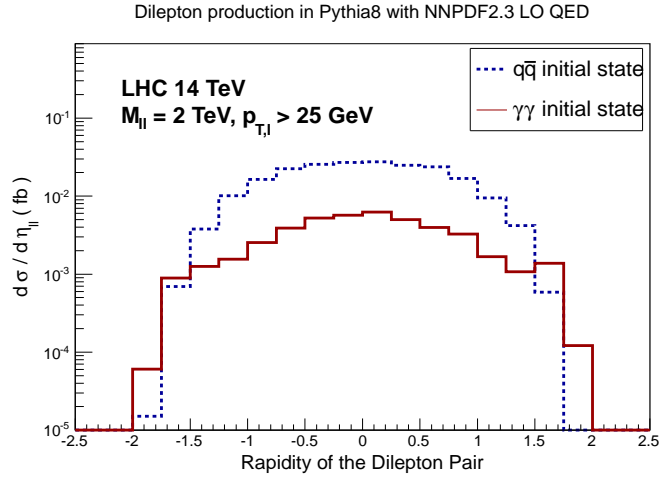}
\caption{\small The rapidity distribution of the dilepton system
at  LHC 14 TeV and for the same kinematical cuts
as in Fig.~\ref{figPythia82}.
\label{figPythia83}}
\end{figure}

In order to disentangle the two  contributions, or to provide
a measurement which is especially sensitive to the photon PDF,
one may look at the rapidity distribution of the dilepton system.
This is shown in Fig.~\ref{figPythia83} for fixed dilepton
invariant mass of 2 TeV, at LHC 14 TeV,  using the same kinematical cuts
as before.
It is clear that for a $q\bar{q}$ initial state, the dilepton system
tends to be produced more centrally (due to the $s$-channel exchange
of the $Z$ boson) while for a $\gamma\gamma$ initial state, the system
is more broadly distributed in rapidity ($t$-channel exchange).
Indeed, for the bins with larger rapidity the contribution from
$\gamma\gamma$ diagrams becomes larger than that of $q\bar{q}$
contributions.

All this suggest that a measurement of the rapidity
distribution of high-mass Drell-Yan pairs, with a cut excluding the central region
to enhance the $\gamma\gamma$ contribution, might be a good way to
isolate and eventually pin down
 the photon contribution to gauge boson production.

\section{Using the NNPDF2.3QED LO sets}

The NNPDF2.3QED LO sets can be obtained from the NNPDF website
\begin{center}
\url{http://nnpdf.hepforge.org/html/nnpdf23qed/nnpdf23qed.html}
\end{center}
together with the corresponding {\tt C/C++} stand-alone code.
They can be used together with the {\tt LHAPDF5.9.0} interface, and 
they will also be available in a future release of {\tt LHAPDF6}.
They are now also  available as an stand-alone internal
PDF set in {\tt Pythia8}. 
For consistency of notation, the NNPDF2.1LO
PDF set (without QED corrections) will henceforth be equivalently
referred to as NNPDF2.3~LO.

{\bf Acknowledgments:}  
We are grateful to T.~Sjostrand and P.~Skands for
  their help with the implementation of the NNPDF2.3 sets in {\tt
    Pythia8}. 

\end{document}